\documentclass[fleqn,usenatbib]{mnras}

\usepackage[T1]{fontenc}
\usepackage{ae,aecompl}
\usepackage{graphicx}
\usepackage{epstopdf}
\usepackage{amsmath}
\usepackage{amssymb}
\usepackage{enumerate}

\title[GRB 170817A - GW 170817]{GRB 170817A as a jet counterpart to gravitational wave trigger GW 170817}
\author[G.P. Lamb and S. Kobayashi]{Gavin P Lamb$^{1,2}$ and Shiho Kobayashi$^{1}$
\\
$^{1}$Astrophysics Research Institute, LJMU, IC2, Liverpool Science Park, 146 Brownlow Hill, Liverpool L3 5RF, 
UK \\
$^{2}$Department of Physics and Astronomy, University of Leicester, University Road, Leicester, LE1 7RH, UK}
\date{Accepted XXX. Received YYY; in original form ZZZ}
\pubyear{2017}
\begin{document}
\label{firstpage}
\pagerange{\pageref{firstpage}--\pageref{lastpage}}
\maketitle

\begin{abstract}
{\it Fermi}/GBM (Gamma-ray Burst Monitor) and INTEGRAL (the International Gamma-ray Astrophysics Laboratory) reported the detection of the $\gamma$-ray counterpart, GRB 170817A, to the LIGO (Light Interferometer Gravitational-wave Observatory)/{\it Virgo} gravitational wave detected binary neutron star merger, GW 170817.
GRB 170817A is likely to have an internal jet or another origin such as cocoon emission, shock-breakout, or a flare from a viscous disc.
In this paper we assume that the $\gamma$-ray emission is caused by energy dissipation within a relativistic jet and we model the afterglow synchrotron emission from a reverse- and forward-shock in the outflow.
We show the afterglow for a low-luminosity $\gamma$-ray burst (GRB) jet with a high Lorentz-factor ($\Gamma$);
a low-$\Gamma$ and low-kinetic energy jet;
a low-$\Gamma$, high kinetic energy jet;
structured jets viewed at an inclination within the jet-half-opening angle;
and an off-axis `typical' GRB jet.
All jet models will produce observable afterglows on various timescales.
{The late-time afterglow from 10-110 days can be fit by a Gaussian structured jet viewed at a moderate inclination, however the GRB is not directly reproduced by this model.}
These jet afterglow models can be used for future GW detected NS merger counterparts with a jet afterglow origin.
\end{abstract}

\section{Introduction}

Short $\gamma$-ray bursts (GRBs) are thought to be due to internal energy dissipation \citep[e.g.][]{1993ApJ...405..278M, 1997ApJ...490...92K, 1998MNRAS.296..275D, 2011ApJ...726...90Z} in an ultra-relativistic jet launched when rapid accretion of material by a compact merger object occurs following a binary neutron star (NS-NS) or neutron star black hole (NS-BH) merger \citep[e.g.][]{1989Natur.340..126E, 1990ApJ...363..218P, 1998ApJ...494L..53K}.
The NS-NS/BH merger is due to the loss of orbital energy and angular momentum via gravitational radiation \citep[e.g.][]{1991ApJ...380L..17P}.
This makes such systems a candidate for gravitational wave (GW) detection by advanced LIGO/{\it Virgo} \citep{2016LRR....19....1A}.
The detection of a GRB in association with a GW signal is key to confirming the neutron star binary merger scenario as the progenitor for short GRBs.

GRB 170817A, with an isotropic equivalent $\gamma$-ray energy $E_\gamma=(4.0\pm0.98)\times10^{46}$ erg at $\sim40$ Mpc, a duration for 90\% of the $\gamma$-ray energy $T_{90}\sim2\pm0.5$ s, and a $\nu F_\nu$ spectral peak energy $E_p=185\pm62$ keV \citep{FermiGCN21506,FermiGCN21528,FermiGRBdetection,INTEGRALGCN,INTEGRALGRBdetection} was detected by {\it Fermi}/GBM and INTEGRAL as a potential electromagnetic (EM) counterpart to the binary NS merger GW 170817 \citep{LVC2017,LVCGBMINTEGRAL} with a delay of $\sim 2$ s from the GW detection to the GRB.
{ From the GW signal, the system is inclined with an angle $0\leq i \leq 36\overset{^\circ}{}$ from the line-of-sight \citep{2017Natur.551...85A}, where the inclination $i$ gives the angle between the rotational axis and the observer}.
{ Using known constraints on $H_0$ the inclination is $3\leq i \leq 23\overset{^\circ}{}$ with the Planck $H_0=67.74\pm0.46$ km s$^{-1}$ Mpc$^{-1}$ \citep{2016A&A...594A..13P}, and $14\leq i \leq 32\overset{^\circ}{}$ using the Type Ia supernova measurements from SHoES $H_0=73.24\pm1.74$ km s$^{-1}$ Mpc$^{-1}$ \citep{2016ApJ...826...56R};
more recently, an inclination of $i=18\pm8\overset{^\circ}{}$ using $H_0$ from the Dark Energy Survey was found by \cite{2017arXiv171203958M}. } 

{ The Swope Supernova Survey detected an optical counterpart (SSS17a) in association with the galaxy NGC4993, $10.9$ hours post-merger \citep{2017arXiv171005452C}.}
The counterpart was consistent with a blue kilo/macro-nova from the dynamical merger ejecta \citep[e.g.][]{2014ApJ...780...31T, 2015MNRAS.446.1115M, 2016AdAst2016E...8T, 2016ApJ...829..110B, 2017arXiv170507084W}.
{See also \citep[][etc]{2017Natur.551...64A, 2017NatAs...1..791C, 2017ApJ...848L..17C, 2017arXiv171005443D, SwiftUVOT, 2017ApJ...849L..19G, 2017arXiv171005434K, 2017ApJ...848L..18N, 2017Natur.551...67P, 2017Sci...358.1574S, 2017Natur.551...75S, 2017PASJ...69..102T, 2017ApJ...848L..27T, 2017ApJ...848L..24V}.}
If GRB 170817A was from internal dissipation within a compact merger jet then the GRB would be accompanied by an afterglow.
In this paper we calculate the expected flux at various frequencies from a forward- and reverse- shock.
We model the afterglow from a low-luminosity GRB jet, a low Lorentz factor ($\Gamma$) jet, structured jets with either a two-component, power-law, or Gaussian structure, and a GRB seen off-axis from a homogeneous jet with typical parameters.

In \S \ref{afterglow} the jet models and parameters used to predict the afterglows are described. In \S \ref{discussion} we discuss the results and their implications for GRB 170817A, and in \S \ref{conc} we give final comments.
\section{Afterglow Prediction}\label{afterglow}
Energy dissipation within an ultra-relativistic jet that results in a GRB will be followed by a broadband afterglow as the jet decelerates in the ambient medium;
depending on the jet parameters, the peak magnitude and timescale at various frequencies can vary significantly.
By assuming that GRB 170817A was from a compact-merger jet viewed either within or outside the jet opening angle we can make reasonable predictions for the expected afterglow.
A forward-shock afterglow is expected to accompany all on-axis GRBs, although a reverse-shock may also be present at early times and typically at low frequencies.

In the following section we calculate the afterglow from forward- and reverse- shocks for a { high-Lorentz factor, low kinetic energy GRB jet} \citep[e.g.][]{1998ApJ...497L..17S,1999ApJ...519L..17S,2000ApJ...542..819K}, and for low-Lorentz factor{ , low and high kinetic energy} jets \citep[e.g.][]{2016ApJ...829..112L}.
We also calculate a forward-shock afterglow for various jet structure models viewed off the central axis, and a homogeneous jet viewed outside the jet half-opening angle \citep[e.g.][]{2017arXiv170603000L}.

\subsection{High-$\Gamma$, Low Kinetic Energy Jet}

Using the isotropic $\gamma$-ray energy reported by {\it Fermi} for GRB 170817A, $E_\gamma = (4.0\pm0.98)\times10^{46}$ erg, and making reasonable assumptions for the afterglow parameters, a prediction can be made for the expected flux at various frequencies.
The typical parameters for a sample of short GRBs are given by \cite{2015ApJ...815..102F} who find that the ambient density is $n \sim (3-15)\times 10^{-3}$ cm$^{-3}$, and the $\gamma$-ray efficiency\footnote{{ The efficiency of the prompt-emission from an internal shock origin is usually given by $\eta\sim f_{\rm dis} \varepsilon_e f_{\rm rad}$ where the fraction of energy dissipated is $f_{\rm dis}\lesssim0.5$, and the fraction of energy radiated is $f_{\rm rad}\sim1$. Using $\varepsilon_e=0.1$, the value of the efficiency should be $\eta\lesssim0.05$. However, the value estimated from an internal shock efficiency can be much higher if we consider the collision of multiple shells with a broad range of Lorentz factors \citep{2001ApJ...551..934K}. The resultant lightcurve would appear smoother and broader for a large number of shells. We base our estimates first on the central observed values of $\eta$ found for short GRBs by \cite{2015ApJ...815..102F} where the range of observed efficiencies is $10^{-3}\lesssim\eta\lesssim0.98$.}} is $0.4 \la \eta \la 0.7$.
As the $\gamma$-ray luminosity of GRB 170817A is well below the typical values for a short GRB, we extend the efficiency range to a lower limit of 0.1;
{ for a jet with an efficiency lower than 0.1, see the discussion at the end of \S \ref{lowG}}.
From the efficiency and $\gamma$-ray energy the jet kinetic energy can be determined, $E_{\rm k}=E_\gamma(1/\eta-1)$;
the jet kinetic energy drives the afterglow.
{ The accelerated particle distribution index for short GRBs is $p=2.43^{+0.36}_{-0.28}$ \citep{2015ApJ...815..102F}, we use $p=2.5$ as our fiducial value.}
Other assumed jet parameters are the jet bulk Lorentz factor, $\Gamma=80$, and the microphysical parameters, $\varepsilon_B=0.01$, and $\varepsilon_e=0.1$.
{ Note that these parameters are assumed throughout unless otherwise stated.}

The duration of the GRB can be used to indicate the width of the relativistic shell, $\Delta_0\sim cT_{90}$ \citep{1997ApJ...490...92K}, where we assume that the GRB is from internal dissipation processes and $c$ is the speed of light.
If the bulk Lorentz factor is below a critical value $\Gamma_c=(3 E_{\rm k}/32\pi n m_p c^2 \Delta_0^3)^{1/8}$, then the reverse shock cannot effectively decelerate the shell;
here $m_p$ is the mass of a proton.
For short GRBs the reverse shock is typically described by the thin shell case.
The shell crossing time for such a reverse shock is $\sim(\Gamma/\Gamma_c)^{-8/3}T_{90}$ and the characteristic frequency for the reverse shock is $\nu_{m,RS}\sim \nu_{m,FS}/\Gamma^2$ \citep{2000ApJ...545..807K}, where subscripts $RS$ and $FS$ indicate reverse- and forward- shocks respectively and $\nu_{m,FS}$ is the forward shock characteristic frequency.
The spectral peak flux at the characteristic frequency is proportional to the number of electrons, the magnetic field, and the bulk Lorentz factor.
The mass in the shell is a factor $\Gamma$ larger than the heated and swept up ambient density of the forward shock region.
The spectral peak flux for the reverse shock is then $F_{\nu,{\rm max},RS}\sim \Gamma F_{\nu,{\rm max},FS}$.
{ The forward- and reverse- shock regions can have a different pre-shock magnetization parameter $\varepsilon_B$, for simplicity we assume that they are the same.}

At low frequencies synchrotron self-absorption becomes important;
for the reverse shock, synchrotron self-absorption will limit the flux more efficiently than for the forward shock because the effective temperature of the electrons in the reverse-shock region is lower by a factor $\sim\Gamma$.
The limiting flux, at a given frequency $\nu$ and observer time $t$, for the reverse shock is \citep[e.g.][]{2000ApJ...542..819K}
\begin{equation}
F_{\nu, {\rm BB}} \sim 2\pi m_p c^2 \Gamma^3 D^{-2} \varepsilon_e t^2 \nu^2 \left(\frac{p-2}{p-1}\right)\left(\frac{e}{\rho}\right){\rm max}\left[\left(\frac{\nu}{\nu_{m,RS}}\right)^{1/2},~1\right],
\label{SSA}
\end{equation}
{ where $e$ is internal energy density and $\rho$ is the mass energy density in the reverse shock region.
At the shock crossing time $(e/\rho) \sim 1$, and  $(e/\rho) \propto t^{-2/7}$ after the shock crossing.}
For the forward shock, the limiting flux is a factor $\Gamma$ larger at the shock crossing time.

{ If the ejecta from the central engine is magnetized, $\varepsilon_B$ in the reverse shock region would be higher than that in the forward shock region.
The higher $\varepsilon_B$ will make the reverse shock peak slightly later and brighter.
At early times and low frequencies, synchrotron self-absorption limits the reverse shock emission. 
As the reverse shock region expands, the emitting surface becomes larger, and the flux limit grows as $F_{\nu,{\rm BB}} \propto t^{1/2}$ \citep[see][for the black body approximation]{2000ApJ...542..819K, 2015ApJ...806..179K} where $\nu<\nu_{m,RS}$. 
When this limit becomes higher than the synchrotron flux  $F_{\nu} \propto \varepsilon_B^{1/3} t^{-1/2}$ \citep{2000ApJ...545..807K}, the reverse shock component peaks. 
Note that the self-absorption limit does not depend on $\varepsilon_B$, but the synchrotron flux does.  
By equalizing the two flux estimates, we find that the peak time and peak flux of the reverse shock emission are scaled as $\varepsilon_B^{1/3}$ and $\varepsilon_B^{1/6}$  respectively.
If $\nu > \nu_{m,RS}$, these scalings are $F_{\nu,{\rm BB}}\propto t^{9/7}$, and $F_\nu \propto \varepsilon^{(p+1)/4} t^{-2}$.
We find the peak time and flux are scaled as $\varepsilon_B^{(p+1)/4}$ and $\varepsilon_B^{(p+1)/10}$ respectively.
For low-$\Gamma$ outflows, synchrotron self absorption is less important and the reverse shock will peak at the time when the shock crosses the shell.
If $\nu_{m,RS}<\nu$ at peak time, then the peak time and flux are proportional to $\varepsilon_B^{(p-1)/4}$ and $\varepsilon_B^{3(p-1)/4}$, \citep[e.g.][]{2005ApJ...628..315Z, 2007ApJ...655..391K}.}
The forward-shock lightcurve will evolve as $t^{-3(p-1)/4}$ after the peak.

A jet viewed on-axis will exhibit a lightcurve break when $\Gamma^{-1}<\theta_j$ \citep{1999ApJ...519L..17S}, where $\theta_j$ is the jet half-opening angle.
As $\Gamma\propto E^{1/8}n^{-1/8}t^{-3/8}$, the break time should occur at 
\begin{equation}
t_j \sim 10 ~E_{{\rm k},50}^{1/3} n_{-2}^{-1/3} (\theta_j/0.31)^{8/3} ~{\rm days},
\label{tjet}
\end{equation}
where subscripts follow the convention $N_x=N/10^x$, $\theta_j$ is in radians and we normalise to a jet with $\theta_j=0.31$ rad, or $\sim 18\overset{^\circ}{}$.
{ Note that for GRB 170817A to be on-axis i.e. within the jet opening angle, the value of $\theta_j$ should be larger than the system inclination.}
For jets where the kinetic energy is $\la10^{48}$ erg, or the half-opening angle is $\la 6\overset{^\circ}{}$, then the jet will break at $\sim 1$ day.
Where the energy is low and the jet is narrow, then the break will occur at $\sim 0.1$ days.
The jet half-opening angle is unknown, { however as the inclination is $\sim18\overset{^\circ}{}$ \citep{2017arXiv171203958M} this can be used to indicate a wide jet if the GRB is observed on-axis.}
The jet-break is not included in the analysis.

\begin{figure}
\includegraphics[width=\columnwidth]{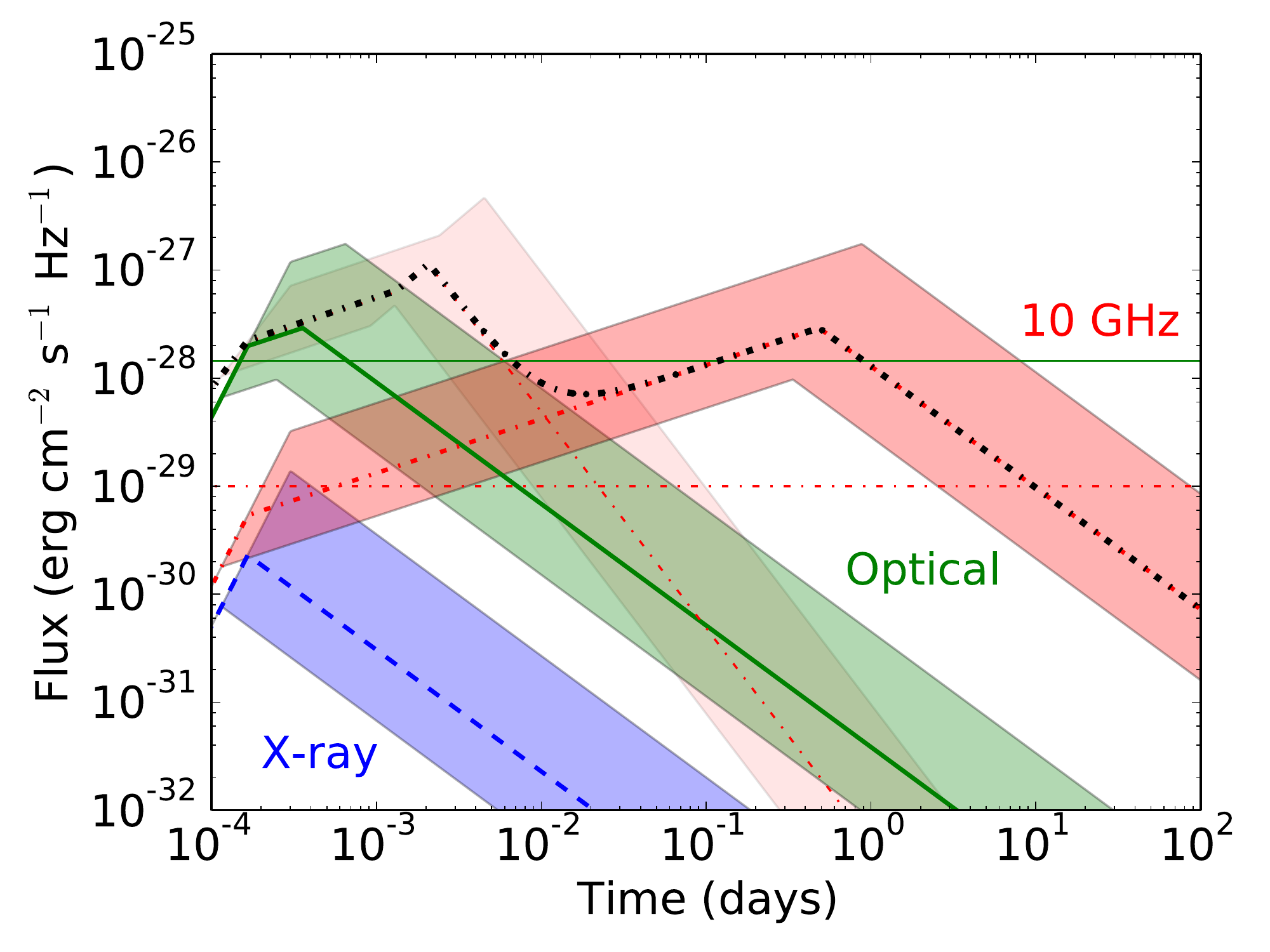}
\caption{Afterglow lightcurves for a jet with an isotropic $\gamma$-ray energy of $4.0\times10^{46}$ erg, a $\gamma$-ray efficiency of $\eta=0.4$, a jet bulk Lorentz factor $\Gamma=80$, in an ambient medium of $n=0.009$ cm$^{-3}$ with microphysical parameters $\varepsilon_B=0.01$ and $\varepsilon_e=0.1$, and a luminosity distance of 40 Mpc. The blue dashed line shows the X-ray afterglow, the green solid line shows the optical afterglow, and the red dashed-dotted line shows the 10 GHz radio afterglow. The shaded regions indicate the lightcurve for an efficiency $0.1 \leq \eta \leq 0.7$. The reverse shock is important at radio frequencies, the 10 GHz reverse shock is shown as a thin dash-dotted red line and faint shaded region for the range of jet energies considered; the forward and reverse shock lightcurve at 10 GHz is shown as a thick black dashed-dotted line. The the red dashed horizontal line indicates the $1 \mu$Jy limit, the green horizontal dashed line indicates $m_{AB}\sim21$ magnitude, and lower-limit of the $y$-axis is the X-ray sensitivity $\sim 0.4 ~\mu$Crab at 4 keV}
\label{1}
\end{figure} 

The afterglow lightcurve for a jet viewed on-axis is shown in Figure \ref{1};
the ambient density is set as the mean of the \cite{2015ApJ...815..102F} sample, $n=0.009$ cm$^{-3}$. 
{ Before the deceleration time, when $\Gamma$ is constant, the forward shock flux and characteristic frequency depend on the ambient density as $[F_{\nu, {\rm max}},~\nu_m]\propto n^{1/2}$.
The deceleration time depends on the number density as $t_{dec}\propto n^{-1/3}$.
After the deceleration time, $\nu_m\propto t^{-3/2}$ and the dependence on the ambient density vanishes.
Where $\nu<\nu_m$ at the deceleration time, the lightcurve will continue to increase as $F_\nu=F_{\nu,{\rm max}}(\nu/\nu_m)^{1/3}$ until $\nu=\nu_m$, the peak here is therefore $F_\nu\propto n^{1/2}$ as $\nu_m$ no-longer depends on $n$}.

Afterglow lightcurves are shown for 10 GHz, optical, and X-ray frequencies.
The shaded regions represent the uncertainty in the $\gamma$-ray efficiency $0.1 \leq \eta \leq 0.7$.
The bold afterglow lines show the lightcurve for a $\gamma$-ray efficiency $\eta=0.4$, where the dashed-dotted red line is 10 GHz, the solid green line is optical ($5\times10^{14}$ Hz), and the dashed blue line is X-ray ($10^{18}$ Hz).
The reverse shock emission is shown as a thin dashed-dotted red line with a faint shaded region;
and the reverse- and forward- shock afterglow at 10 GHz assuming the mean efficiency is shown as a thick black dashed-dotted line.
The forward shock dominates emission for optical and X-ray frequencies.
As a reference, the horizontal dashed-dotted line shows $1~\mu$Jy, horizontal solid line shows $m_{\rm AB}=21$, and the approximate {\it Swift}/XRT (X-Ray Telescope) limit is given by the lower-limit of the $y$-axis at $10^{-32}$ erg cm$^{-2}$ s$^{-1}$ Hz$^{-1}$.

\begin{figure}
\includegraphics[width=\columnwidth]{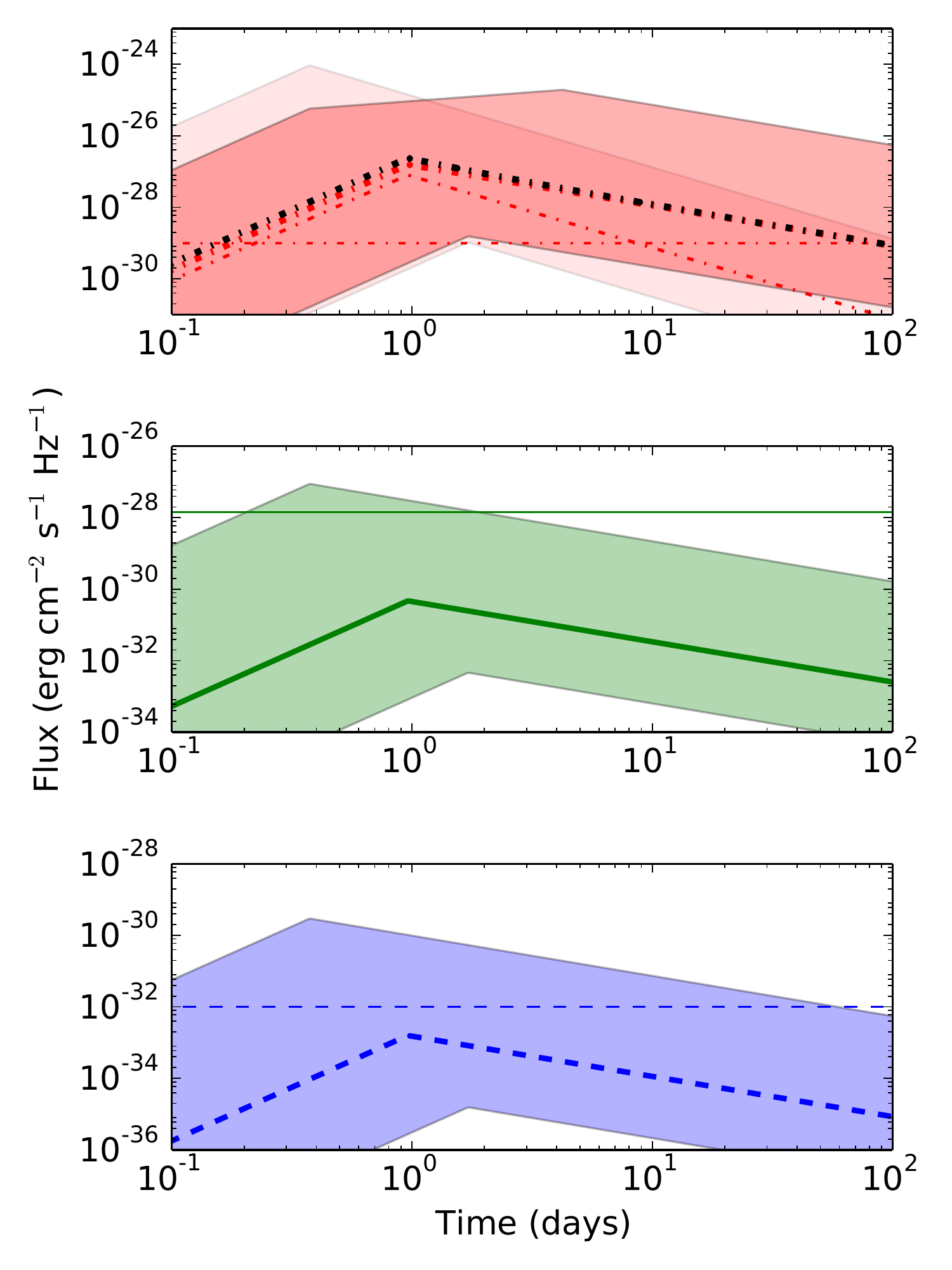}
\caption{Afterglow from a low-$\Gamma$ jet with an isotropic $\gamma$-ray energy of $4.0\times10^{46}$ erg, a $\gamma$-ray efficiency of $0.001\leq\eta\leq0.7$ and a luminosity distance 40 Mpc. The jet bulk Lorentz factor is estimated from the delay time as $2.2 \la \Gamma \la 10.0$, all other parameters are as Figure \ref{1}. The lines show the afterglow for a jet with $\Gamma\sim3.9$, the shaded regions indicate the uncertainty in the kinetic energy and the Lorentz factor. Colours are as for figure \ref{1}. {\bf Top} panel: 10 GHz emission where the thin dashed-dotted line and faint shaded region indicate the reverse-shock; the thick dashed-dotted line and shaded region indicate the forward-shock; the sum of reverse- and forward- shock light-curves is shown as a black dashed-dotted line. The red horizontal dashed line indicates the 1 $\mu$Jy limit. {\bf Middle} panel: optical afterglow. The green solid line shows the optical magnitude 21. {\bf Bottom} panel: X-ray afterglow. The blue horizontal dashed line is $\sim 0.4$ $\mu$Crab at $\sim4$ keV}
\label{2}
\end{figure}

\subsection{Low-$\Gamma$ Jets}\label{lowG}

The minimum radius at which the prompt $\gamma$-ray photons can be emitted is the photospheric radius, where the outflow becomes optically thin.
The photospheric radius is given by
\begin{equation}
R_p=\left[\frac{\sigma_T E_{\rm k}}{4\pi m_p c^2\Gamma}\right]^{1/2}\sim 1.9\times10^{13} E_{{\rm k},50}^{1/2} \Gamma_{1}^{-1/2}~{\rm cm},
\label{Rp}
\end{equation}
where $\sigma_T$ is the Thomson cross-section.

Considering the relatively high $E_p$ despite the low $L_\gamma$ we assume that the prompt $\gamma$-ray photons are emitted near the photosphere.
The observed delay time between the GW signal and the GRB is equivalent to the travel time for a constant Lorentz factor flow to a radial distance equivalent to the photospheric radius, $\Delta t\sim R_p/2\Gamma^2c$.
The bulk Lorentz factor is then
\begin{equation}
\Gamma = \left[ \frac{\left(\sigma_T E_{\rm k}\right)^{1/2}}{4 \Delta t c^2 \left( \pi m_p \right)^{1/2}}\right]^{2/5} \sim 12~ E_{{\rm k},50}^{1/5} \left(\frac{\Delta t}{2~{\rm s}}\right)^{-2/5} ,
\label{gamma}
\end{equation}
where $\Delta t$ is the measured delay time.

The prompt $\gamma$-ray emission is predicted to be suppressed for a jet with a low Lorentz factor{ , the higher energy emission will be suppressed due to pair production and the total energy in the photons reduced due to adiabatic cooling before decoupling from the expanding plasma at the photosphere,}\footnote{ { This suppression results in the fraction of energy radiated being $f_{\rm rad}<1$ while the assumed value for $\varepsilon_e$ remains unchanged.}} \citep[e.g.][]{2014ApJ...782....5H,2016ApJ...829..112L}.
{ GRB 170817A had a thermal component \citep{FermiGRBdetection} that would be expected from photospheric emission \citep[e.g.][]{2006ApJ...642..995P}.}
To reflect { the possible prompt suppression} we extend the lower limit of the $\gamma$-ray efficiency range\footnote{{ Where the efficiency is high, the jet kinetic energy will be low and suppression of dissipated energy within a low-$\Gamma$ outflow reduced \citep[see][]{2016ApJ...829..112L}. Such low-energy, low-luminosity, and low-$\Gamma$ jets may form a distinct population \citep[e.g.][]{2016arXiv160603043S}}}.
The Lorentz factor for a jet with $0.001 \leq \eta \leq 0.7$, and the observed $E_\gamma$, from equation \ref{gamma}, is $10.0\ga\Gamma\ga2.2$.
The afterglow lightcurves from low-$\Gamma$ jets are shown in Figure \ref{2};
we use an efficiency of $\eta=0.1$ for the lightcurve.
The shaded region indicates the afterglow for the limits of the efficiency.

The low-$\Gamma$ value for the outflow gives a relatively long deceleration time ($t_{\rm dec}$) for the jet, where $t_{\rm dec}\propto \Gamma^{-8/3}$.
The reverse shock will cross the shell at $\sim 0.4-1.7$ days for $10\ga\Gamma\ga2.2$ respectively.
At radio frequencies the reverse-shock emission will dominate over the forward-shock lightcurve at $t_{\rm dec}$ for $\Gamma\ga5$.
This will result in a brightening of the lightcurve before the forward-shock peak due to the reverse-shock.
The reverse-shock is only important at early times and for the upper limits of the parameter space;
the reverse-shock is shown for the 10 GHz lightcurves in Figure \ref{2}.

\begin{figure}
\includegraphics[width=\columnwidth]{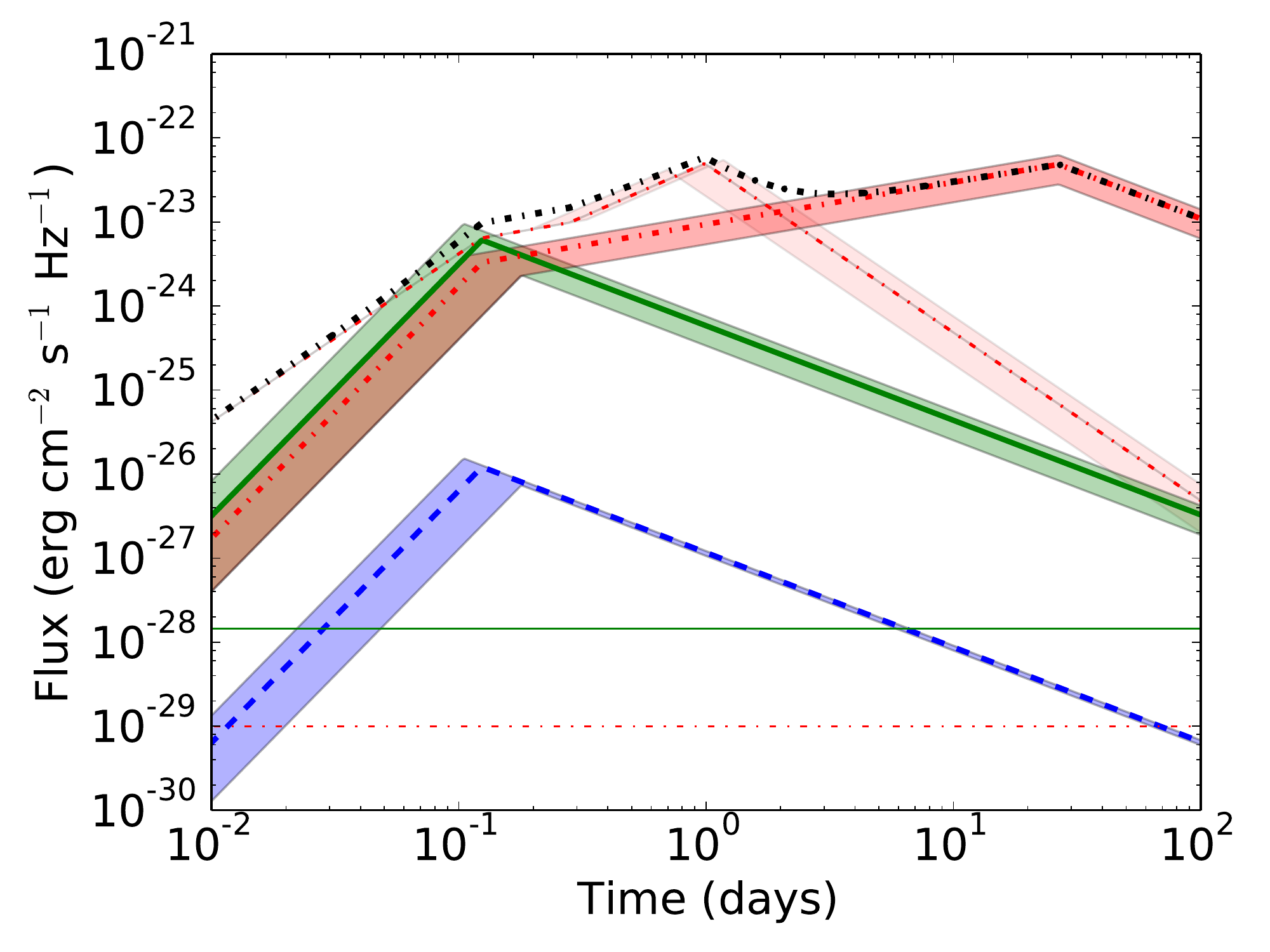}
\caption{Afterglow from a low-$\Gamma$ jet with a jet kinetic energy of $10^{52}$ erg,  and a luminosity distance 40 Mpc. The jet bulk Lorentz factor is estimated from the delay time as $\Gamma \sim 30$. Shaded regions represent the range of ambient densities $(3 \la n \la 15)\times10^{-3}$ cm$^{-3}$, all other parameters are as Figure \ref{1}. The reverse-shock at 10 GHz is shown as a thin dash-dotted red line and faint shaded region. Colours are as for Figure \ref{1}. The green horizontal solid line is optical $m_{AB}=21$, and the red horizontal dashed-dotted line indicates the 1 $\mu$Jy limit}
\label{3}
\end{figure}

The level of suppression of the prompt emission is unknown;
if all jets from binary neutron star mergers produce jets with a similar kinetic energy \cite[e.g.][]{2017PhRvD..95j1303S}, then the afterglow would appear brighter than a low-luminosity jet afterglow with a typical $\eta$ value.
Using a jet kinetic energy of $E_{\rm k}=10^{52}$ erg, the bulk Lorentz factor from equation \ref{gamma}, would be $\Gamma \sim30$ and the prompt emission significantly suppressed \citep[e.g.][]{2016ApJ...829..112L}.
{ The prompt efficiency for such a jet would be very low, $\eta\sim10^{-6}$, where the observed GRB had energy equivalent to GRB 170817A.}
The afterglow for such a jet is shown in Figure \ref{3};
as the jet kinetic energy is fixed, here the limits of the shaded regions represent the uncertainty on the ambient medium number density, $n\sim (3-15)\times10^{-3}$ cm$^{-3}$.
A reverse-shock is apparent at 10 GHz, peaking at $\sim2$ days with a flux $\sim10$ Jy;
the reverse-shock is shown in the figure as a thin red dashed-dotted line with the associated uncertainty in the ambient number density.
A black dashed-dotted line indicates the sum of the 10 GHz lightcurve from the forward- and reverse- shocks.

\subsection{Structured Jet}\label{struct}

\begin{figure*}
\includegraphics[width=\textwidth]{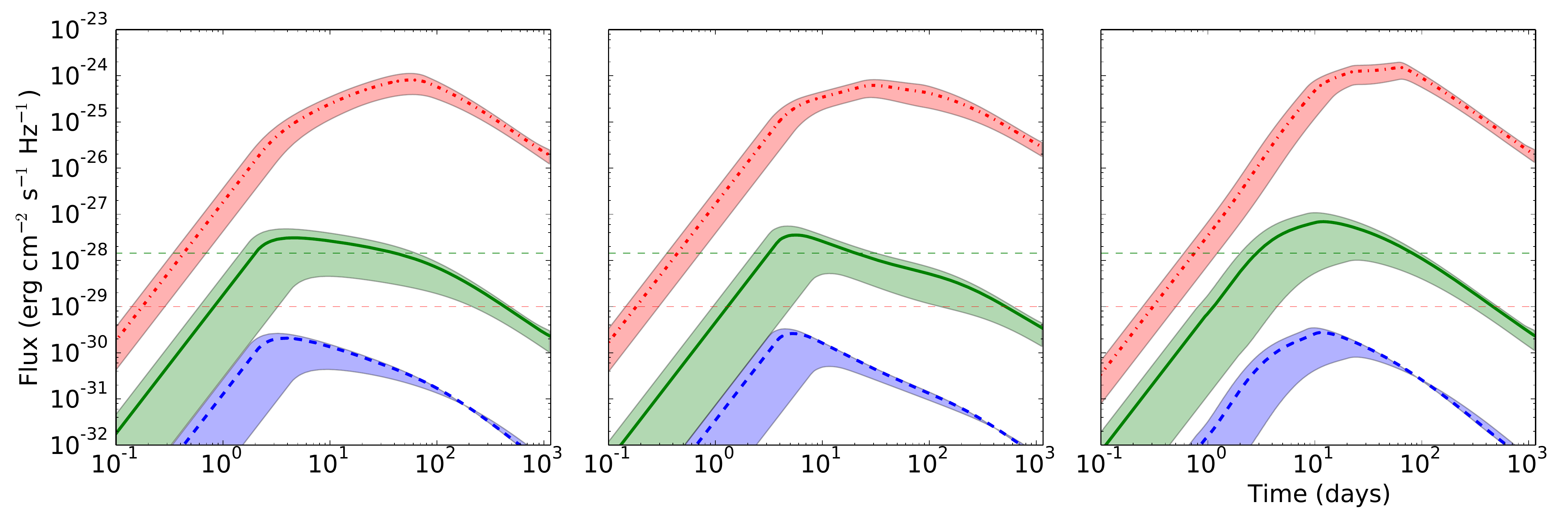}
\caption{Afterglows from jets with structure; jet core parameters are $E_{\rm{ iso}}=10^{52}$ erg, $\eta=0.4$, $\Gamma=80$, and $\theta_c=6\overset{^\circ}{}$, all other parameters are as previously used. The jet structure extends to 25$\overset{^\circ}{}$ in each case. {\bf Left}: Gaussian structure, a Gaussian function on $E$ and $\Gamma$ with angle from the centre. Jet inclined to the observer at $18\overset{^\circ}{.}5$. {\bf Middle}: Power-law structure with a decay index outside of the core of $k=-2$. Jet inclined to the observer at $25\overset{^\circ}{.}5$. {\bf Right}: Two-component structure, where the second component has 5\% of the core parameters. Jet inclined to the observer at $11\overset{^\circ}{}$.}
\label{LCs}
\end{figure*}

GRBs are usually assumed to have a homogeneous, or `top-hat', structure i.e. the energy and Lorentz factor are uniform in a jet cross-section and the jet has a sharp edge defined by the jet half-opening angle.
However, jets may have some intrinsic structure either due to the formation and acceleration processes or as a result of jet breakout from merger ejecta.
Here we use the structured jet models from \cite{2017arXiv170603000L};
see also \cite{2017arXiv171000275X} for a similar analysis { or \cite{2017arXiv170807008J} and \cite{2018MNRAS.473L.121K} for discussion of the prompt emission from a structured jet}.
For each of the three models used the total isotropic equivalent jet core energy is fixed at $10^{52}$ erg, and the core extends to an angle of 6$\overset{^\circ}{}$ from the central axis.
The jet parameters, $E$ and $\Gamma$, vary according to the model:
for a two-component jet, $E$ and $\Gamma$ are at 5\% of the core values between $(6-25)\overset{^\circ}{}$;
for a power-law jet, $E$ and $\Gamma$ vary with angle outside the core following a power-law index -2;
and for a Gaussian structured jet the parameters $E$ and $\Gamma$ depend on angle following a Gaussian function from $(0-25)\overset{^\circ}{}$.
The detected prompt emission in a 50-300 keV band is determined for each jet model at observation angles from $(0-25)\overset{^\circ}{}$ and a distance 40 Mpc.
The observation angle values are selected for each jet structure where the detected prompt photon flux is comparable to the observed {\it Fermi}/GBM and INTEGRAL.
{The prompt emission from each jet component is calculated considering the angle to the line-of-sight, and the dissipation and photospheric radius in each case.
The flux at the detector is determined by considering the photon arrival times and the emission duration.}
The afterglow from each model for the determined inclination is then generated following the method in \cite{2017arXiv170603000L}.

The Gaussian jet model, shown in Figure \ref{LCs} left panel, has an inclination of $18\overset{^\circ}{.}5$.
For the power-law jet model, shown in Figure \ref{LCs} central panel, the inclination angle is $25\overset{^\circ}{.}5$.
For the two-component model, shown in Figure \ref{LCs} right panel, the inclination angle is 11$\overset{^\circ}{}$;
{ note that for the two-component model the $\gamma$-ray emission is that seen off-axis from the core jet region, the wider sheath component has a low-$\Gamma$ value such that the prompt emission is fully suppressed.}
In the figure the afterglow at 10 GHz is shown in red with a dashed-dotted line, optical is shown in green with a solid line, and X-ray is shown in blue and with a dashed line.
The shaded region represents the uncertainty in the ambient medium number density, with the line indicating the afterglow for the mean $n=0.009$ cm$^{-3}$.

For each model the first break in the lightcurve is due to the deceleration time for the jet component inclined towards the observer, i.e. the jet-component at the inclination angle.
At radio frequencies, the lightcurve will peak when the characteristic frequency crosses the observation band, $\nu_m=\nu$.
At optical and X-ray frequencies, and at radio frequencies for the two-component jet, a late-time excess or a shallow decay is due to the off-axis emission from the bright core of the jet.
Any late-time break in the lightcurve is due to the edge of the jet becoming visible i.e. the jet-break, equation \ref{tjet}.

For the structured jet models the photon flux at the detector from the prompt emission approximates, without fine-tuning, the observed parameters:
for the Gaussian jet the prompt fluence is $\sim 3.8\times10^{-7}$ erg cm$^{-2}$;
for the power-law jet the prompt fluence is $\sim7\times 10^{-7}$ erg cm$^{-2}$;
and for the two-component jet the prompt fluence is $\sim2.1\times 10^{-7}$ erg cm$^{-2}$.
The {\it Fermi}/GBM measured fluence is $(2.8\pm0.2)\times 10^{-7}$ erg cm$^{-2}$ \citep{FermiGRBdetection, LVCGBMINTEGRAL}.
The difference in fluence between the jet models and the observed value is due to the choice of numerical resolution.
The fluence for each jet model was calculated in $0\overset{^\circ}{.}5$ steps from $0 - 28\overset{^\circ}{}$ and the inclination for the jet determined by the angle for which the fluence was closest to the observed value.
{The observed spectral shape, or peak value, was not calculated in this estimation.}

\subsection{Off-Axis Afterglow}

\begin{figure}
\centering
\includegraphics[width=\columnwidth]{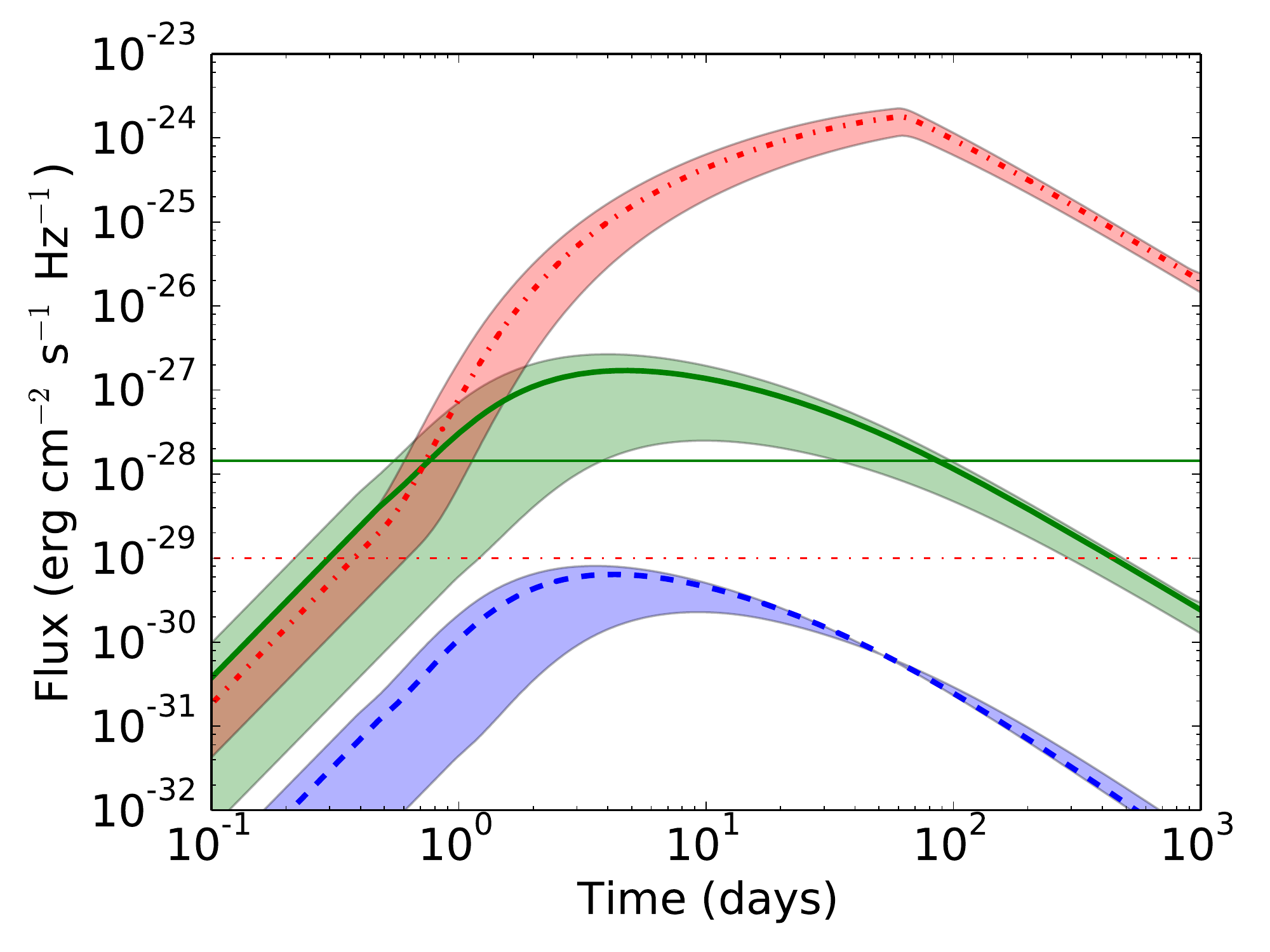}
\caption{Off-axis afterglow from a homogeneous jet with $E_{\rm{iso}}=10^{52}$ erg, $\eta=0.4$, $\Gamma=80$, and a half-opening angle $\theta_j=6\overset{^\circ}{}$. The observed $\gamma$-ray fluence in the 50-300 keV band is $2.1\times10^{-7}$ erg cm$^{-2}$; the inclination from the jet central axis is $11\overset{^\circ}{}$ and the ambient density is in the range $0.003 \leq n \leq 0.015$ cm$^{-3}$.}
\label{OA}
\end{figure}

The $T_{90}$ duration of GRB 170817A is longer than the typical value of $\sim0.6$ s \citep{2012ApJ...750...88Z}, although still within the usual period for short GRB classification $\la 2$ s.
The delay time between the GW signal and the detected prompt emission, and the duration and low-luminosity of the $\gamma$-rays could be due to the jet inclination to the line-of-sight;
where for an off-axis observer the time until emission and the duration are lengthened from that for an on-axis observer by the relativistic Doppler factor, $t\propto\delta^{-1}$ where $t$ is the observed time, $\delta=[\Gamma(1-\beta\cos \theta_{\rm obs})]^{-1}$ is the Doppler factor and $\beta$ the velocity as a fraction of $c$, and the observed fluence is $\propto\delta^{3}$ \citep[e.g.][]{2001ApJ...554L.163I}.
The off-axis prompt emission will also appear to be brighter in X-rays \citep[e.g.][]{2002ApJ...571L..31Y}.

If the jet is inclined in such a way that the observer's line-of-sight is outside of the jet edge i.e. $\theta_{\rm obs}>\theta_j$, then the prompt and afterglow emission will be delayed and suppressed when compared to that seen by an on-axis observer i.e. $\theta_{\rm obs}\rightarrow 0$.
In considering an observer at various angles from the jet central axis, we use the method in \cite{2017arXiv170603000L} which includes the jet geometry and emission surface to determine the inclination at which the prompt $\gamma$-ray photons have a similar fluence\footnote{We do not change any of the prompt energy parameters from the model in \cite{2017arXiv170603000L} except the total isotropic energy, efficiency, and bulk-Lorentz factor, where we use $E=10^{52}$ erg, $\eta=0.4$ and $\Gamma=80$ instead of $E=2\times10^{52}$ erg, $\eta=0.1$ and $\Gamma=100$. This maintains consistency with earlier scenarios and avoids fine-tuning.}.
At an inclination of $11\overset{^\circ}{}$ for a jet with $\theta_j=6\overset{^\circ}{}$, $E_{\rm iso}=10^{52}$ erg, an efficiency $\eta=0.4$, and a $\Gamma=80$, the simplest estimate of the fluence in a $T_{90}$ period from our model is $2.1\times10^{-7}$ erg cm$^{-2}$.
The corresponding afterglow in an ambient medium $0.003 \leq n \leq 0.015$ cm$^{-3}$ is shown in Figure \ref{OA} where the colours are as previous figures.
Note that as $\nu_a<\nu<\nu_m$ at the deceleration time for the 10 GHz lightcurve, then the synchrotron self-absorption frequency $0.25 \la \nu_a \la 0.75$ GHz at this time will not affect the lightcurve \citep{1999ApJ...519L..17S}.

Given an observed $E_p\sim185$ keV, and the inclination, jet half-opening angle and $\Gamma$ used, the `on-axis' spectral peak energy would be a few MeV.
Short GRBs with a spectral peak of a few MeV include GRB 061006, 070714, and 090510;
where the $E_p=[955\pm267,~2150\pm1113,~\rm{and}~8370\pm760]$ keV respectively \citep[e.g.][]{2012ApJ...750...88Z,2016CRPhy..17..617P}.
All of these GRBs have high luminosities for short GRBs, where $L_\gamma>10^{52}$ erg s$^{-1}$.
{ The high on-axis $E_p$ value applies to the two-component jet discussed in \S \ref{struct}, where the wider sheath component has no detectable $\gamma$-ray emission and only contributes to the afterglow lightcurve.} 

\section{Discussion}\label{discussion}

By assuming that the observed GRB is from a compact merger jet we have shown the expected afterglow lightcurves for various jet models.
If GRB 170817A was a low-luminosity GRB viewed on-axis, the afterglow in X-ray and optical would peak within seconds of the GRB.
A reverse-shock in the radio, typically fainter than $\la$1 mJy at 10 GHz, { may} be visible peaking on a timescale of minutes;
this will be followed by the radio forward-shock afterglow peak with flux $\la$ 0.1 mJy at $\sim 1$ day i.e. Figure \ref{1}.
The predicted optical afterglow is fainter than $m_{AB}\la19$,
and the X-ray afterglow is detectable by {\it Swift}/XRT but will fade rapidly.
The X-ray afterglow will peak within seconds and typically last $\sim 15$ minutes before becoming too faint for {\it Swift}/XRT, where we assume an X-ray limit of $>10^{-32}$ erg cm$^{-2}$ s$^{-1}$ Hz$^{-1}$.
Such a fast and faint transient would be challenging to detect.

By considering the delay time from GW signal to GRB, constraints can be put on the jet bulk Lorentz factor, if the jet is inclined within the half-opening angle i.e. on-axis.
The energy dissipated will decouple from the jet when the optical depth becomes unity, at the photospheric radius.
By using an assumed $\gamma$-ray efficiency, the jet kinetic energy can be estimated and from this and the delay time a value for $\Gamma$ found.
The bulk Lorentz factor found using an efficiency $0.001\leq\eta\leq0.7$ is $10.0\geq\Gamma\geq2.2$ respectively.
This is consistent with the low-$\Gamma$ jet model of \cite{2016ApJ...829..112L} where the prompt emission is expected to be significantly suppressed.
The forward shock afterglow from such a jet is shown in Figure \ref{2};
the afterglow peak in all bands is $\la1$ day and optical and X-ray are faint.
Radio emission at 10 GHz is typically $\la1$ mJy, and would be detectable for $\ga1-100$ days.

If the $\gamma$-ray efficiency is very low i.e. the jet kinetic energy is $E_{\rm k}>>E_\gamma$ then the derived bulk Lorentz factor, using $E_{\rm k}=10^{52}$ erg, is $\Gamma\sim30$.
This value is consistent with the low-$\Gamma$ jet model, predicting suppression of the prompt emission resulting in a low-luminosity GRB.
The afterglow for such a jet is shown in Figure \ref{3};
the peak afterglow is typically a few hours after the GRB at optical and X-ray frequencies.
Radio, optical, and X-ray emissions are bright in all cases.
The 10 GHz afterglow remains at the $\sim 1$ Jy level for $\sim 10-1000$ days, while optical and X-ray fade rapidly.

A jet with extended structure may naturally produce low-luminosity GRBs at wider angles where the jet energetics are lower.
By following the structured jet models of \cite{2017arXiv170603000L}, we show the expected afterglow from a jet with these models where the observed $\gamma$-ray flux is equivalent to the detected {\it Fermi} value.
The afterglows from a Gaussian jet viewed at $i=18\overset{^\circ}{.}5$, a power-law jet viewed at $i=25\overset{^\circ}{.}5$, and a two-component jet viewed at an inclination $i=11\overset{^\circ}{}$ are shown in Figure \ref{LCs}.
Radio, optical and X-ray emissions are bright in all cases with optical and X-ray lightcurves peaking $\sim 3-100$ days, and 10 GHz at $\sim20-100$ days at the 0.1-1 Jy level.
Various features are distinct for each jet model:
the Gaussian jet has an early peak with a shallow { rise or} decline in optical and X-ray emission for $\sim100$ days before breaking to a more rapid decline.
In addition the radio typically peaks at the break.
{ For an observer at a wider inclination, the afterglow lightcurve will show a slow rise from a few days to a peak at $\gtrsim100$ days at all frequencies \citep[e.g.][]{2017arXiv170603000L}.}
The power-law jet has a sharp early peak at optical and X-ray frequencies whilst the 10 GHz afterglow has a later peak with a slower increase in flux after the deceleration time.
Finally the two-component jet has a softer peak and shows a slight rebrightening at late times, especially at radio frequencies, before a rapid decline.

An observer at an inclination just higher than the jet's half-opening angle will see the relativistically beamed prompt and afterglow emission at a later time and at a lower frequency and intensity.
The observed delay in the prompt emission, and the low-luminosity can be explained by the jet inclination;
the afterglow in such a case would be similarly delayed and fainter.
We show the afterglow for an observer at $11\overset{^\circ}{}$ from the jet central axis, where the jet has a half-opening angle $\theta_j=6\overset{^\circ}{}$, an isotropic equivalent blast energy $10^{52}$ erg, a $\gamma$-ray efficiency of $\eta=0.4$, and $\Gamma=80$.
The X-ray afterglow, at $\sim4$ keV, rises slowly to a peak flux $\la10^{-30}$ erg cm$^{-2}$ s$^{-1}$ Hz$^{-1}$ at $\sim 30$ days;
optical afterglow has a similar rise index and peak time with a $m_{AB}\la16$;
while the 10 GHz afterglow has a steeper rise rate, breaking to a soft peak from 70 days, the 10 GHz afterglow is brighter than 1 $\mu$Jy from $\ga1-2$ days and peaks at $\sim 1$ Jy.

A neutron star binary merger is expected to produce a kilo/macro-nova that will peak with a thermal spectrum at optical to near-infrared frequencies during the first 10 days \citep[e.g.][]{2014ApJ...780...31T, 2015MNRAS.446.1115M, 2016AdAst2016E...8T,  2016ApJ...829..110B, 2017arXiv170507084W}.
For the structured or off-axis jet afterglows, the optical emission may peak on a similar timescale to the expected kilo/macro-nova.
However X-ray and radio emission will reveal the afterglow in such a case.
Non-detections by X-ray and/or radio searches for an afterglow from GRB 170817A at { early, $<10$ days,} times can be used to rule out the various structured, and high kinetic energy with low-$\Gamma$ jet scenarios { presented here}.

The prompt emission for GRB 170817A was fit by an exponential cut-off power-law, the Comptonization spectrum model \citep[e.g.][]{2016A&A...588A.135Y}, with a $\nu F_\nu$ spectral peak energy at $E_p\sim185\pm62$ keV and an index $\alpha \sim -0.62\pm0.40$ \citep{LVCGBMINTEGRAL, FermiGCN21506, FermiGCN21528, FermiGRBdetection, INTEGRALGRBdetection}.
Due to the sparsity of high-energy photons, the requirement for an ultra-relativistic bulk Lorentz factor is relaxed.
Additionally, with this $E_p$ and low luminosity, the GRB does not fit on the $E_p-L_\gamma$ correlation for all GRBs \cite[e.g.][]{2010PASJ...62.1495Y, 2012ApJ...750...88Z}.
{ A structured jet where the photospheric emission is treated more precisely could explain the GRB \citep{2018arXiv180101410M}, or} the $\gamma$-rays could be { due to inefficient particle acceleration,} wider angle Comptonized emission,{ or scattered jet internal prompt emission \citep{2015ApJ...809L...8K, 2017arXiv171100243K}.
Alternatively the detected $\gamma$-ray flux may not have been from a jet but a more isotropic outflow \citep[e.g.][]{2018MNRAS.474L...7S};
a cocoon or shock-breakout \citep{2006ApJ...652..482P, 2017ApJ...834...28N, 2017MNRAS.471.1652L, 2018MNRAS.473..576G}, or a flare due to fragmentation of a viscous disc \citep{2006ApJ...636L..29P}.}

\section{Conclusions}\label{conc}

We have modelled the afterglow from various jet dynamical scenarios given the observed $\gamma$-ray flux detected by {\it Fermi} and INTEGRAL for GRB 170817A in association with the GW signal GW 170817.
Four scenarios were considered: (i) an on-axis low-luminosity GRB with typical high Lorentz factor; (ii) low-$\Gamma$ jets viewed on-axis; (iii) jets with extended structure where the prompt emission would have an energy similar to that observed; (iv) and an off-axis jet where the prompt emission is geometrically corrected to give the observed $\gamma$-ray fluence.
In all cases an afterglow is expected on various timescales and with a range of peak fluxes.
Where the kinetic energy is typical for a GRB jet, the afterglow for either a low-$\Gamma$ jet or from a structured jet where the prompt $\gamma$-ray emission is suppressed or low, will result in a bright afterglow, easily detectable at all frequencies.
If GRB 170817A is from within a relativistic jet then the jet must be either:
\begin{itemize}
\item A low energy jet with either a low- or high- $\Gamma$, and a high $\gamma$-ray efficiency $\eta\ga0.4$
\item A GRB jet viewed off-axis
\end{itemize}
If the jet is the first of these, then a large population of low-luminosity, low-energy jets from neutron star mergers could exist resulting in a high GW detection rate for neutron star mergers.

\begin{figure}
\centering
\includegraphics[width=\columnwidth]{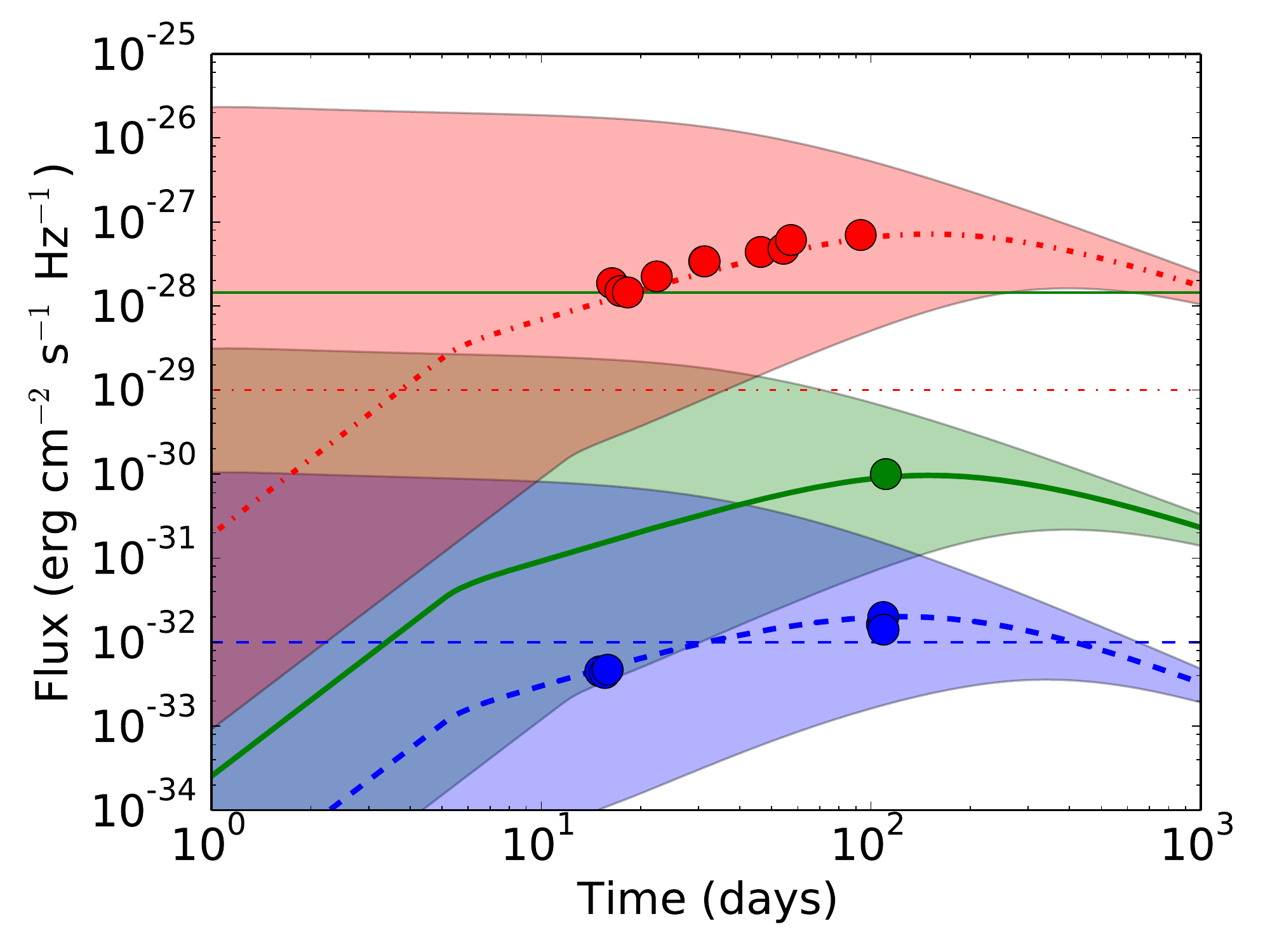}
\caption{Afterglow from a modified Gaussian structured jet with parameters tuned to recreate the observed radio, optical and X-ray observations if viewed at $20\overset{^\circ}{}$. The afterglow range indicates the lightcurve for an inclination $10\leq i\leq26\overset{^\circ}{}$.
X-ray at 1 keV is shown in blue, optical with green, and 3 GHz in red.
Markers indicate observations from Hallinan et al. (2017), Haggard et al. (2017), Lyman et al. (2018), Margutti et al. (2017), Mooley et al. (2017), and Ruan et al. (2017); errorbars are typically smaller than the markers and not included.}
\label{new}
\end{figure}

\subsection{An Evolving Afterglow}
X-ray and radio counterparts have been initially reported from $\sim9-18$ days post-merger \citep{VLAGCN21815, VLA, 2017ApJ...848L..20M, VLAGCN21814,2017Natur.551...71T}.
Radio counterparts are expected from the merger ejecta at late times \cite[e.g.][]{2016ApJ...831..190H}.
{ However, the X-ray and radio observations from $\sim10-100$ days \citep{2017ApJ...848L..25H, VLA, 2017ApJ...848L..20M, 2017arXiv171111573M, 2017arXiv171202809R} and recent optical data \citep{NATPAP} are consistent with a Gaussian structured jet.
One phenomenological fit is for an observer at $\sim 20\overset{^\circ}{}$, and with the parameters used in \S \ref{struct} tuned  \citep[e.g.][]{NATPAP, 2018arXiv180103531M}, the jet energy structure is a modified Gaussian profile, $e^{-\theta^2/\theta_c^2}$ and a Gaussian profile for the Lorentz factor, $e^{-\theta^2/2\theta_c^2}$.
The parameters for the afterglow shown in Figure \ref{new} are $E_{\rm k}=10^{52}$ erg, $\Gamma=80$ for the jet core with an angle $\theta_c=4\overset{^\circ}{.}5$, microphysical parameters $\varepsilon_e=0.01$ and $\varepsilon_B=0.01$, $p=2.1$, and $n=10^{-3}$ cm$^{-3}$;
where the range indicates an observer between $10\leq i\leq26\overset{^\circ}{}$ \citep{2017arXiv171203958M} and the thick lines indicate $20\overset{^\circ}{}$.
The GRB emission is not directly reproduced by this model, however the contribution from scattered prompt emission of the jet core \citep{2017arXiv171100243K} or other higher latitude effects have not been considered.
Alternatively a jet-cocoon structure can explain the observed afterglow or a choked-jet cocoon \citep{2017arXiv171203237L, 2017arXiv171111573M}.}

{The afterglow models presented here can be used with future EM jet-counterparts to GW detected NS mergers.
For a Gaussian structured jet, the rising broadband emission of the afterglow from $\sim10$ days depends on the inclination and the jet parameters, whereas for a cocoon model it should be fairly consistent for a wide range of observation angles.
Failed GRB afterglows, or other jet structures could be revealed by further GW-EM detections.}

\section*{Acknowledgements}
The authors thank the anonymous referee, Iain A. Steele, Phil James, Dan Hoak, David Bersier, and Hendrik van Eerten for useful comments.
This research was supported by STFC grants.
GPL was partially supported by IAU and RAS grants.

\bsp
\label{lastpage}
\end{document}